\documentclass[]{spie}  %>>> use for US letter paper
\usepackage{amsmath,amsfonts,amssymb}
\usepackage{graphicx}
\usepackage[colorlinks=true, allcolors=blue]{hyperref}

\usepackage{xparse}
\usepackage{hyperref}

\title{First Demonstration of Kernel Phase Interferometry on JWST/MIRI: Prospects for Future Planet Searches Around Post Main Sequence Stars}

\author[a]{Chelsea Adelman}
\author[b]{Steph Sallum}
\author[c]{Matthew De Furio}
\author[d]{Josh Eisner}
\affil[a]{University of California, Irvine, Irvine, CA, USA}
\affil[b]{University of California, Santa Cruz, Santa Cruz, CA, USA}
\affil[c]{University of Texas at Austin, Austin, TX, USA}
\affil[d]{University of Arizona, Tucson, AZ, USA}

\authorinfo{Further author information: (Send correspondence to Chelsea Adelman)\\Chelsea Adelman: E-mail: cladelma@uci.edu\\  Steph Sallum: E-mail: ssallum@ucsc.edu\\ Matthew De Furio: E-mail: defurio@utexas.edu\\ Josh Eisner: E-mail: jeisner@arizona.edu}

\begin{document} 
\maketitle

\begin{abstract}
Kernel phase interferometry (KPI) is a post-processing technique that treats a conventional telescope as an interferometer by accurately modeling a telescope pupil as an array of virtual subapertures. KPI provides angular resolution within the diffraction limit by eliminating instrumental phase errors to first order. It has been successfully demonstrated to boost angular resolution on both space- and ground-based observatories, and is especially useful for enhancing space telescopes, as their diameters are smaller than the largest ground-based facilities. Here we present the first demonstration of KPI on \textit{JWST}/MIRI data at 7.7 microns, 10 microns, and 15 microns. We generate contrast curves for 16 white dwarfs from the MIRI Exoplanets Orbiting White dwarfs (MEOW) Survey, finding significantly deeper contrast at small angular separations compared to traditional imaging with \textit{JWST}/MIRI, down to within $\lambda$/D. Additionally, we use our KPI setup to successfully recover four known companions orbiting white dwarfs and brown dwarfs. This analysis shows that at these wavelengths KPI can uniquely access the orbital parameter space where inward-migrating post-main-sequence giant exoplanets are now thought to exist. We discuss the prospects for applying KPI to a larger sample of white dwarfs observed with \textit{JWST}, increasing the volume of directly imaged close-in post-main-sequence exoplanets. 
\end{abstract}


\section{INTRODUCTION}\label{sec:intro} 
Kernel phase interferometry (KPI) is a post processing technique that has been successful in resolving known companions, and in detecting unknown companions unattainable with standard imaging, on both space-based \cite{browndwarf_HST, jens_niriss, me_steph} and ground-based observatories\cite{palomar, Jens_VLT, keck_kpi}. By modeling a conventional telescope as a densely-packed interferometer, KPI accesses angular separations that extend within the diffraction limit of the telescope \cite{Fizeau, StephAndy, Ireland2013}. It is best used for moderate contrast imaging, and requires high Strehl ($\gtrsim$0.85)\cite{StephAndy}. KPI has also been successfully demonstrated on \textit{JWST}/NIRISS \cite{jens_niriss}, but has not yet been implemented on other \textit{JWST} instruments. The combination of \textit{JWST's} stability, high Strehl, and long wavelength would make long-wavelength \textit{JWST} KPI useful for detecting colder companions than those accessible from the ground or with \textit{HST} KPI.

Here we present the first demonstration of KPI on \textit{JWST}/MIRI. KPI's moderate contrast makes it well suited for constraining the occurrence rates of post-main-sequence (post-MS) planets around white dwarfs (WDs), which has recently been demonstrated as an exciting \textit{JWST} science application \cite{WDPlanetCandidate, SED_transit_conf_and_MEOW, MEOW_detection, poulsen_MIRI, DIexoplanets}. \textit{JWST's} sensitivity also makes KPI a viable technique for companion searches around faint brown dwarfs (BDs). As a proof of concept, we utilize archival data of white dwarfs and brown dwarfs for KPI analysis.  We create achievable contrast curves for 16 of the 17 targets in the MIRI Exoplanets Orbiting White dwarfs (MEOW) survey (PID 4403). We also attempt to recover known companions orbiting two BDs, 014656.66+423410.0\cite{discovery_0416} and J171104.60+350036.8\cite{companion_1711} (referred to as W0416 and W1711, respectively, hereafter), which were observed as part of a \textit{JWST} cycle one program (PID 2124). Additionally, we attempt to recover the two first directly imaged exoplanets orbiting WDs, WD 1202-232 and WD 2105-82\cite{DIexoplanets}, which were discovered via another \textit{JWST} cycle one program (PID 1911).

In this manuscript, we discuss KPI in greater detail in Section \ref{sec:KPI} and the interferometric model creation and parameters in Section \ref{sec:methods}. We then describe the data and the reduction techniques in detail in Section \ref{sec:datared}. We present the results in Section \ref{sec:results} and discuss their implications as well as future plans in Section \ref{sec:disc}. Lastly, we conclude in Section \ref{sec:conc}.

\section{Kernel Phase Interferometry}\label{sec:KPI}
Kernel phase interferometry treats a conventional telescope like an interferometer \cite{Fizeau}. The amplitudes and phases of complex visibilities are calculated via the Fourier transform (FT) of the images. The filled aperture is modeled as a redundant interferometric array with a linear relationship between instrumental phases (which are assumed to be small; $\varphi << 1$) and Fourier phases. Instrumental errors can be eliminated to the first order by using a linear combination of the Fourier phases, called kernel phases (KPs).

The following equations describe the the kernel phase calculations. Equation 1 describes the linear model that connects the pupil plane phases and the Fourier phases.

\boldmath{}
\begin{equation}
\label{eq:start}
\Phi = \mathrm{R^{-1}} \cdot \mathrm{A} \cdot \varphi + \Phi_0
\end{equation}
Here $\Phi$ is a vector of measured complex visibility phases, and $\mathrm{R}$ is a square diagonal matrix consisting of redundancy values of the baselines. \boldmath{$\mathrm{A}$} is a matrix representing the pairs of subapertures contributing to each spatial frequency in the FT, and \unboldmath{$\varphi$} represents the instrumental phases associated with the subapertures in the pupil plane. Multiplying both sides by \boldmath{$\mathrm{R}$} gives

\boldmath{}
\begin{equation}
\label{eq:middle}
\mathrm{R} \cdot \Phi = \mathrm{A} \cdot \varphi + \mathrm{R} \cdot \Phi_0.
\end{equation}
To eliminate the instrumental errors, the left null space of $\mathrm{A}$, called $\mathrm{K}$, is formulated. Left multiplying $\mathrm{K}$ on both sides of Equation 2 eliminates the instrumental term and leaves the intrinsic phase information of the target. This is shown in Equation 3.

\begin{equation}
\label{eq:end}
\mathrm{K} \cdot \mathrm{R} \cdot \Phi = \mathrm{K} \cdot \mathrm{R} \cdot \Phi_0
\end{equation}
\unboldmath{}

\section{KPI Pupil Model Generation}\label{sec:methods}
\begin{figure} [ht]
   \begin{center}
   \begin{tabular}{c} %% tabular useful for creating an array of images 
   \includegraphics[height=5cm]{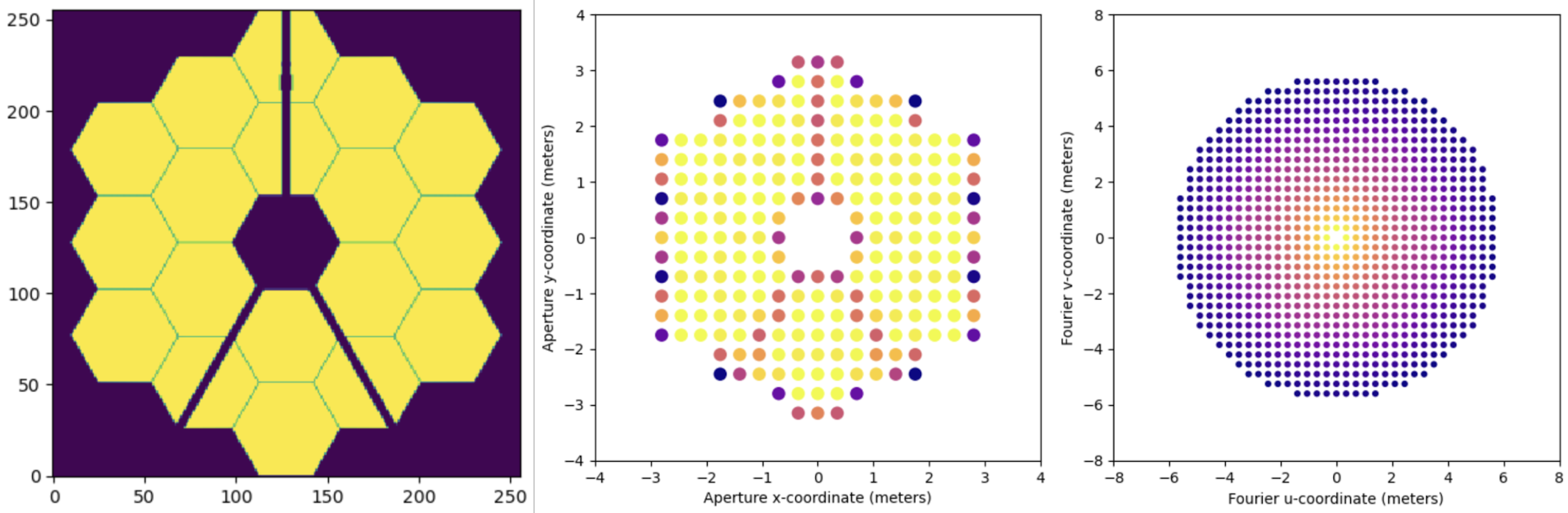}
   \end{tabular}
   \end{center}
   \caption[pupil] 
%>>>> use \label inside caption to get Fig. number with \ref{}
   { \label{fig:pupil} 
\textbf{Left:} A model image of the \textit{JWST}/MIRI pupil is created. \textbf{Middle:} The pupil image is used to create a model interferometric array. The sub-apertures are regularly spaced and each one with another forms a baseline, which we can use to sample a specific spatial frequency. The colorscale represents the varying transmission $0.70\leq t \leq 1$. \textbf{Right:} The resulting discrete (u,v) coverage from the baselines created between each subaperture in the middle image. The colorscale corresponds to Fourier amplitude, decreasing from the center to the outer edges.}
   \end{figure}

Constructing the \boldmath{$\mathrm{K}$}\unboldmath{} matrix requires a pupil model, which we generate using the Python package now known as the Space Telescope Point Spread Function (\texttt{STPSF}\footnote{https://github.com/spacetelescope/stpsf}) formerly known as \texttt{webbpsf}. This package calculates simulated PSFs for \textit{JWST}. We use the \texttt{STPSF} \textit{JWST}/MIRI pupil model with \texttt{XARA}\footnote{https://github.com/fmartinache/xara/}, a python package designed for KPI calculations. \texttt{XARA} can be used to create ``grey" or ``binary" models. Here, binary means transmission is either 0\% or 100\% for all subpertures throughout all parts of the pupil. Grey models account for diffraction that occurs around the spiders, secondary mirror, mirror segment gaps, and edges of the primary mirror, where transmission will be between 0\% and 100\%. Using \texttt{XARA}, we create a grey interferometric model and set the transmission cutoff (the transmission value where subapertures would be excluded) to be 0.70, previously used with KPI on \textit{JWST}/NIRISS \cite{jens_niriss}. This model can be found in Figure \ref{fig:pupil}. 

We define the pupil model pitch, $s$, or the spacing between each subaperture, as 0.35m. The pitch is chosen by visual inspection of the (u,v) points where sampling takes place in the observed Fourier-transformed images (see Section \ref{sec:KPI}). It is chosen to result in sampling dense enough to resolve phase variations in the FT. Since the pitch determines the lowest sampled spatial frequency, when using KPI to constrain an object's morphology, the effective field of view (FoV) is set by the pitch and wavelength, where the FoV $\le \frac{\lambda}{s}$. As described in Section \ref{sec:datared}, it is also good practice to use this effective field of view to set the minimum cropped image size. We use this pupil model to perform a KPI analysis of WDs and BDs observed in various filters (see Section \ref{sec:observations}), and discuss strategies for optimizing its parameters in Section \ref{sec:disc}.

\section{Observations}\label{sec:observations}
The data in this work come from the publicly available Barbara A. Mikulski Archive for Space Telescopes (MAST) and they were collected using \textit{JWST}'s Mid-Infrared Instrument (MIRI). We use three different datasets from three different \textit{JWST} programs (PIDs: 4403, 1911, and 2124) to perform a MIRI KPI proof of concept. We analyze 18 WDs and two BDs. The observations for each dataset are described below.

\textbf{\textit{The MEOW Survey (PID 4403)}}: We reduce data taken as part of the MEOW Survey with the goal of generating contrast curves. The data were collected from September 9, 2023 to May 4, 2024 and comprise 17 WDs observed with \textit{JWST}/MIRI. All targets were observed in F770W, F1800W, and F2100W, corresponding to 7.7 $\mu$m, 18 $\mu$m, and 21 $\mu$m. The exposure times for each target in the F770W, F1800W, and F2100W filters are 55.5s, 277.504s, and 710.412s, respectively. One target, G 107-70, is in a tight binary, so we chose to exclude it from our analysis. We focus on the other 16 targets, specifically those observed in the F770W filter.

\textbf{\textit{Brown Dwarfs: W0146 and W1711 (PID 2124)}:} We reduce the W0146 and W1711 data with the goal of recovering known, close-in companions. The data for W1711 and W0146 were collected on May 31, 2023 and September 21, 2022, respectively. W0146 has a known companion orbiting at $\sim150$ mas, $\sim190^\circ$ E of N with a contrast of $\sim1$ mag, and W1711 has a known companion orbiting at $\sim701$ mas, $\sim341^\circ$ E of N with a contrast of $\sim0.074$ mag \cite{BD_mathew}. Both targets are observed with F1000W, F1280W, and F1800W filters corresponding to 10 $\mu$m, 12.8 $\mu$m, and 18 $\mu$m, respectively. Exposure times for F1000W, F1280W, and F1800W images are 38.85s, 44.4s, and 55.5s, respectively for W0146. Exposure times for F1000W, F1280W, and F1800W images are 27.75s, 38.85s, and 55.5s, respectively for W1711. We focus specifically on the data collected with the F1000W filter.

\textbf{\textit{White Dwarfs: WD 1202-232 and WD 2105-82 (PID 1911)}:} We analyzzed observation of WD 1202-232 and WD 2105-82 to recover known wider separation companions. The data were collected for WD 1202-232 and WD 2105-82 on February 9, 2023 and April 21, 2023, respectively. WD 1202-232 has a known companion orbiting at $\sim1.11$$^{\prime\prime}$, $\sim114^\circ$ E of N with a contrast of $\sim4.45$ mag, and WD 2105-82 has a known companion orbiting at $\sim2.14$$^{\prime\prime}$, $\sim200^\circ$ E of N with a contrast of $\sim3.83$ mag. Both targets were observed in F560W, F770W, F1500W, and F2100W, corresponding to 5.6 $\mu$m, 7.7 $\mu$m, 15 $\mu$m, and 21 $\mu$m respectively. The exposure times for F560W, F770W, F1500W, and F2100W data are 255.304s, 277.504s, 8413.92s, and 1309.816s respectively for WD 1202-232. The exposure times for F560W, F770W, F1500W, and F2100W images are 233.104s, 233.104s, 12088.08s, and 6016.288s respectively for WD 2105-82. We specifically focus on the F1500W data for these two targets.

\section{Data Reduction}\label{sec:datared}
We first use aperture photometry to perform background subtraction for each target. To do this we use a Python package called \texttt{photutils}. The median of an annulus around each target is calculated to measure local background noise, then subtracted from the image. Each image is then cropped. As described in Section \ref{sec:methods}, given the pupil sampling pitch, the KPIs would be sensitive to spatial scales as large as $\sim$4.5$^{\prime\prime}$x4.5$^{\prime\prime}$, $\sim$5.8$^{\prime\prime}$x5.8$^{\prime\prime}$, and $\sim$8.8$^{\prime\prime}$x8.8$^{\prime\prime}$ at 7.7 $\mu$m, 10 $\mu$m, and 15 $\mu$m, respectively. Each MEOW target (observed at 7.7 $\mu$m) is cropped to 40x40 pixels, corresponding to a FoV of 4.4$^{\prime\prime}$x4.4$^{\prime\prime}$. The BD data (obtained at 10 $\mu$m) are cropped to 50x50 pixels, corresponding to a FoV of 5.5$^{\prime\prime}$x5.5$^{\prime\prime}$. Both of the images with WDs (observed at 15 $\mu$m) are cropped more aggressively because of faint background sources near the science target. WD 2105-82 is cropped to 64x64 pixels, corresponding to a FoV of 7$^{\prime\prime}$x7$^{\prime\prime}$, and WD 1202-232 is cropped to 30x30 pixels, corresponding to a FoV of 3.3$^{\prime\prime}$x3.3$^{\prime\prime}$. The downside to cropping aggressively is losing the sampling of some of the shorter baselines, or lower spatial frequencies, but because the separations of these companion are small compared to the lowest spatial frequencies, this did not inhibit recovering the signal.

      \begin{figure} [ht]
   \begin{center}
   \begin{tabular}{c} %% tabular useful for creating an array of images 
   \includegraphics[height=7cm]{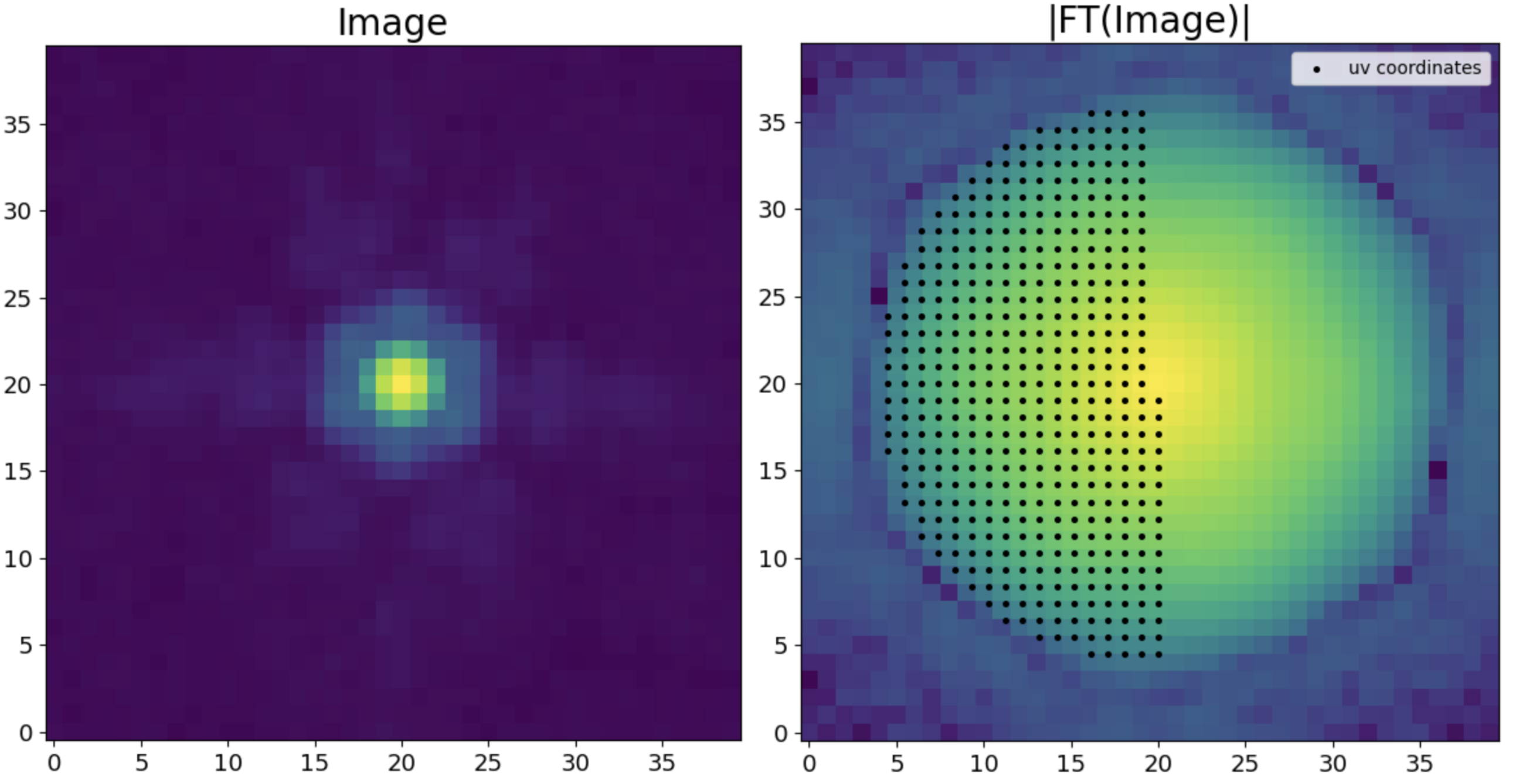}
   \end{tabular}
   \end{center}
   \caption[FT] 
%>>>> use \label inside caption to get Fig. number with \ref{}
   { \label{fig:FT} 
Example reduced and Fourier-transformed images, showing a MEOW target, WD 0839-327. \textbf{Left:} The image of the WD which is cropped, bad pixel corrected, and multiplied with a super-Gaussian window to suppress pixels far from the PSF center. \textbf{Right:} The Fourier transform is taken and sampled at a grid of regularly-spaced uv coordinates, where phases are extracted and then projected into KPs.}
   \end{figure} 

The image is then corrected for bad pixels by comparing each pixel to a 10 x 10 surrounding box. A bad pixel is flagged and replaced with the median of its unflagged neighbors if its value exceeds the box's average by more than $7\sigma$. We then apply a super-Gaussian window to the image to eliminate noisy pixels far from the PSF center. Without this suppression, these noisy pixels can create correlations in the FT. Most of the targets have a super-Gaussian of order four and HWHM defined as $0.5 \frac{\lambda}{s}$ converted to pixels ($\sim21$, $\sim27$, and $\sim40$ pixels for 7.7 $\mu$m, 10 $\mu$m, and 15 $\mu$m, respectively). We are aggressive with the windowing of W0146, with a HWHM of 20 pixels, as noisy pixels far from the PSF were mimicking a signal. Then using \texttt{XARA} and the pupil model, we take the FT of the images and sample the FT at (u,v) coordinates to extract KPs, as shown in Figure \ref{fig:FT}.
We assign conservative error bars by measuring the standard deviation of all of the final KPs and assigning it as a uniform error to every KP. 

The data come from observing programs that were not ideal for performing good calibration because each target is observed once in each filter, and no reference stars were observed. So, for this proof of concept, all calculated KPs are raw and uncalibrated. Typically for calibration we use either angular differential imaging \cite{ADI}\cite{KDI} if we have images observed at two roll angles, or reference star differential imaging (RDI)\cite{Jens_VLT} if unresolved objects were observed close in time to the science target. Interferometric calibration using RDI on \textit{JWST} requires calibrators to be observed as close as possible on the detector and with peak counts as close as possible \cite{steph_AMI}. One of the next steps is to carefully calibrate the KPs for optimization of companion detectability, discussed further in Section \ref{sec:disc}.

We next use \texttt{XARA} functions to create a set of KPs for binary models of various combinations of separations, position angles, and contrasts, creating a three dimensional grid of model KPs for all parameter combinations. The best fit binary model is found by minimizing the $\chi^2$ calculated from the model KPs and the data. If no significant signal is detected, we then use the same grid of companion fits to generate contrast curves. For these, $\chi^2$ values are averaged across the position angles to create a grid of $\chi^2$ as a function of separation and contrast. We use intervals of $\chi^2$ values between the null model, with a separation of 0 mas, and the companion models to determine the detectability confidence levels. We create the final contrast curves by choosing contrasts with $\Delta \chi^2$ of 25, corresponding to being distinguishable from the null model at $5\sigma$ \cite{StephAndy}.

\section{Results}\label{sec:results}
\begin{figure} [ht]
   \begin{center}
   \begin{tabular}{c} %% tabular useful for creating an array of images 
   \includegraphics[height=14cm]{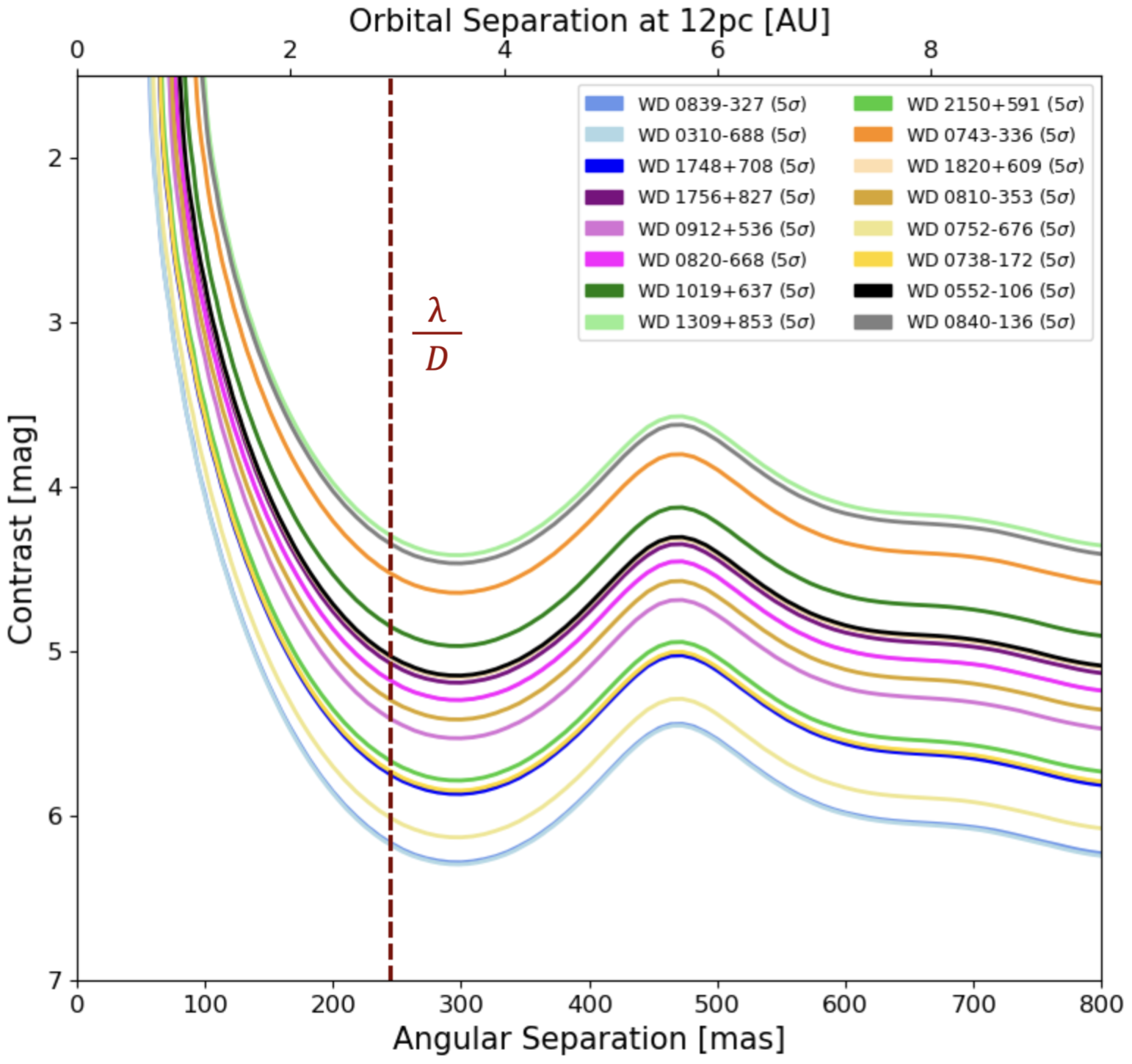}
   \end{tabular}
   \end{center}
   \caption[contrast] 
%>>>> use \label inside caption to get Fig. number with \ref{}
   { \label{fig:contrast} 
Solid curves show $5\sigma$ contrast limits for 16 WDs in the MEOW dataset. The top x-axis shows the orbital separation at 12 pc, the average distance of the 16 WDs, and the bottom x-axis shows the angular separation in mas. The dashed vertical maroon line shows $\lambda/\mathrm{D}$ for 7.7 $\mu$m.}
   \end{figure}

Here we present the contrast curves and companion fitting results. We create $5 \sigma$ contrast curves for 16 of the 17 targets in the MEOW dataset, shown in Figure \ref{fig:contrast}. We achieve contrasts of $5-6$ mag for most MEOW targets within $\lambda/\mathrm{D}$ ($<250$ mas or $<3$AU at the average MEOW target distance of $\sim12$pc away). Standard imaging can achieve contrast of roughly 6 mag at $\sim580$ mas for 5.8 $\mu$m, and deeper contrasts at $>600$mas \cite{poulsen_MIRI} (assuming similar performance, this translates to $\sim780$ mas scaled for a wavelength 7.7 $\mu$m). These results demonstrate that MIRI KPI provides a boost in angular resolution, decreasing the inner working angle by a factor of $\sim4$.

\begin{table}
\begin{center}
\begin{tabular}{|l|cc|cc|cc|}
    \hline
    ID & Sep$_{\mathrm{DI}}$ & Sep$_{\mathrm{KPI}}$  & PA$_{\mathrm{DI}}$ & PA$_{\mathrm{KPI}}$ & Contrast$_{\mathrm{DI}}$ & Contrast$_{\mathrm{KPI}}$ \\
   & [mas] & [mas] & [deg] & [deg] & [mag] & [mag] \\
\hline 
    J014656.66+423410.0 & $149.9^{+26.4}_{-7.70}$ & $190.13^{+1.55}_{-1.55}$ & $273.25^{+3.86}_{-2.23}$ & $310.35^{+1.61}_{-1.61}$ & $1.05^{+0.18}_{-0.15}$ & $0.059^{+0.017}_{-0.017}$  \\
    J171104.60+350036.8 & $701.60^{+9.90}_{-6.60}$ & $696.55^{+12.4}_{-9.30}$ & $341.65^{+2.23}_{-0.62}$ & $336.31^{+0.31}_{-0.31}$ & $0.74^{+0.05}_{-0.05}$ & $0.725^{+0.066}_{-0.053}$  \\      
    WD1202-232 & $1110^{+40}_{-40}$ & $1166^{+24}_{-167}$ & $114.0^{+2.00}_{-2.00}$ & $112.90^{+3.41}_{-3.41}$ & $4.45^{+1.41}_{-1.41}$ & $4.51^{+0.093}_{-0.403}$  \\   
    WD2105-82 & $2140^{+20}_{-20}$ & $2149^{+51}_{-53}$ & $200.40^{+0.40}_{-0.40}$ & $196.51^{+2.28}_{-2.28}$ & $3.83^{+1.01}_{-1.01}$ & $3.95^{+0.372}_{-0.372}$  \\
\hline

\end{tabular}
\label{tab:target_parameters}
\end{center}
\caption{Here we report the separation, position angle, and contrast for the known companions found with direct imaging (DI) and recovered with KPI, as indicated in the subscripts in the table header. The reported DI parameters of the two companions orbiting the BDs \cite{BD_mathew}, and the reported DI parameters of the WD companions \cite{DIexoplanets} are indicated with the subscript 'DI'. KPI and DI result in comparable fits for the wider separation companions, but there are discrepancies for W0416. As discussed in Sections \ref{sec:results} and \ref{sec:disc}, calibrating the KPs may help with recovering such close-in signals.}
\end{table}

\begin{figure} [ht]
   \begin{center}
   \begin{tabular}{c} %% tabular useful for creating an array of images 
   \includegraphics[height=14cm]{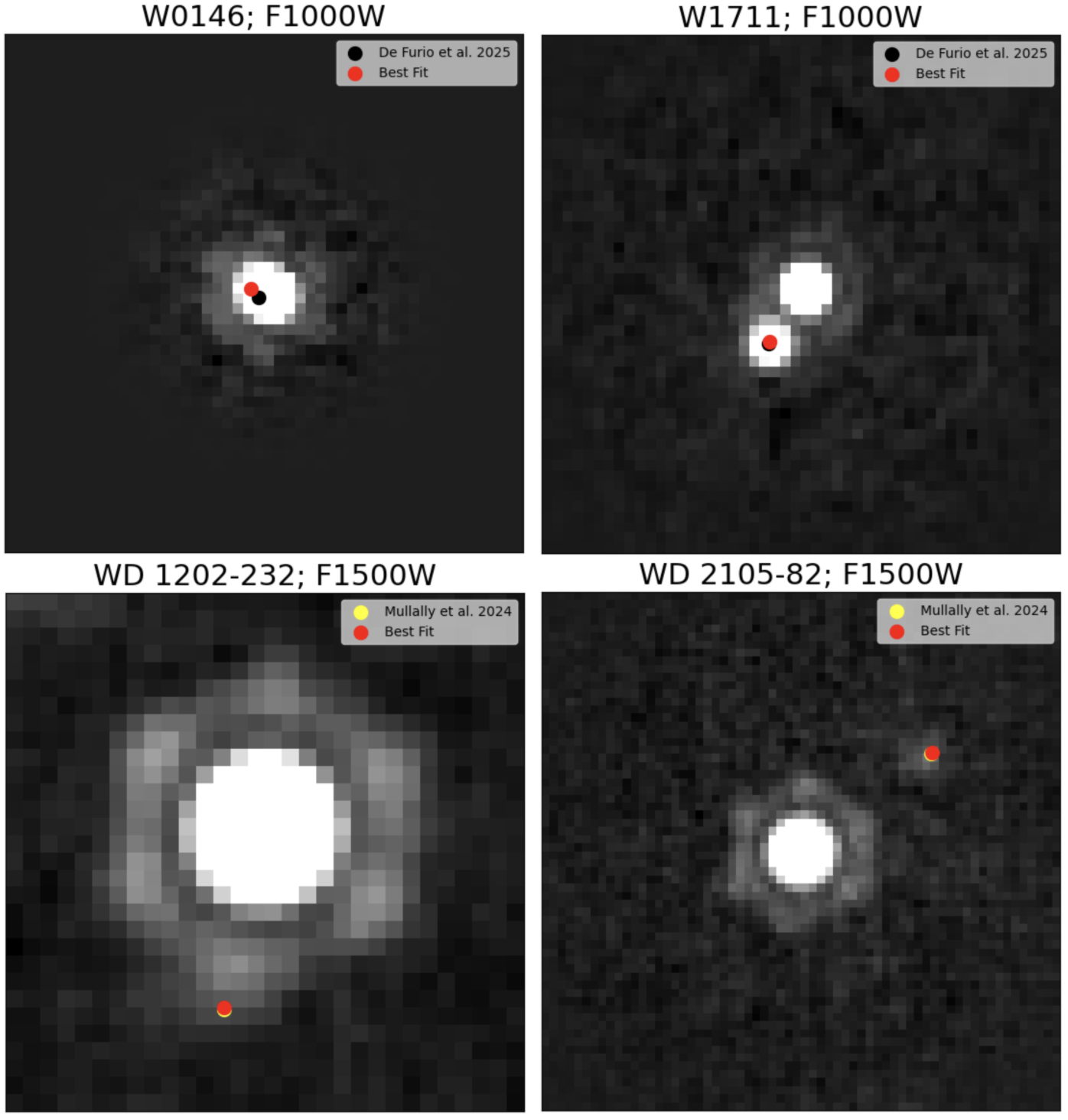}
   \end{tabular}
   \end{center}
   \caption[correlation] 
%>>>> use \label inside caption to get Fig. number with \ref{}
   { \label{fig:comp_images}
Each panel shows one MIRI image of each target with a known, directly-imaged companion. The top two panels are BDs with known companions, W0146 and W1711, and the bottom two panels are WDs, WD 1202-232 and WD 2105-82, with known companions. The black and yellow dots represent the published separations and position angles of the companions as measured by direct imaging, and the red dots indicate the KPI companion parameters. For three of the four targets the KPI and DI fits' error bars overlap, and for W0146 there are larger systematic offsets. We discuss these results in Sections \ref{sec:results} and \ref{sec:disc}.
}
   \end{figure}

All KPI recovered parameters and DI reported parameters for the four companion recoveries can be found in Table \ref{tab:target_parameters}. Figure \ref{fig:comp_images} shows the images and DI/KPI companions positions for the two BDs (W0146 and W1711) and two WDs (WD1202-232 and WD2105-82). As shown in Table \ref{tab:target_parameters}, all DI and KPI companion parameters for W1711 have $1\sigma$ error bars that overlapped. For WD1202-232 and WD2105-82, recovered separations and contrasts have overlapping $1\sigma$ error bars, while the position angles differ by slightly more than $1\sigma$. The differences between the KPI and DI best fit parameters for W0146 are significantly larger than the parameter error bars, at 40.23 mas, 37.1 degrees, and 0.991 mag in separation, position angle, and contrast, respectively. These discrepancies could be caused by systematic errors present in the uncalibrated KPs, which would bias the companion fits and also lead to underestimated parameter error bars (since we assume Gaussian KP errors). We generally expect that kernel phase calibration may cause the KPI outputs to converge to similar values as those reported for the DI analysis of the BD companions \cite{BD_mathew} and WD companions \cite{DIexoplanets}.

\begin{figure} [ht]
   \begin{center}
   \begin{tabular}{c} %% tabular useful for creating an array of images 
   \includegraphics[height=14cm]{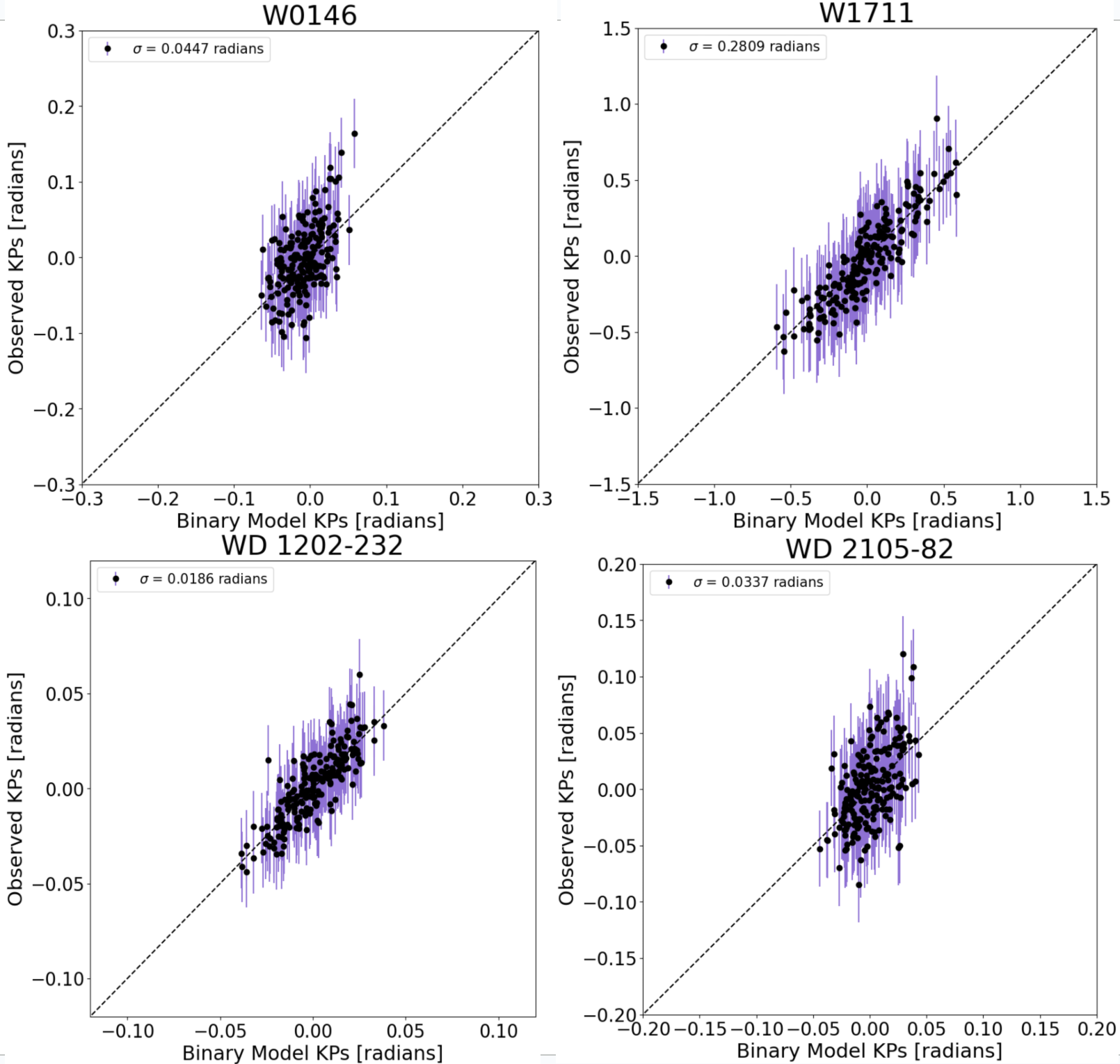}
   \end{tabular}
   \end{center}
   \caption[correlation] 
%>>>> use \label inside caption to get Fig. number with \ref{}
   { \label{fig:correlation} 
The observed KPs are plotted against the model KPs for the best binary fit to the four objects with known companion signals. The top two panels correspond to the two BDs, W0146 and W1711, and the bottom two panels correspond to the two WDs, WD1202-232 and WD2105-82. If a companion signal were present, the data would fall along the dashed black line, indicating a good fit and a one-to-one correlation. The data in the upper right and lower left panels show a strong correlation with the best-fit binary model, and the upper left and lower right show a weaker correlation. These weaker correlations may be due to higher order systematic errors that have not been removed since the KP data are uncalibrated (Sections \ref{sec:results} and \ref{sec:disc}).}
   \end{figure}

The KPs associated with the best fit binary parameters are compared with the KP data in Figure \ref{fig:correlation}. If a companion were present with high SNR, the data would lie along the black dashed line, indicating a one-to-one correlation. We see that W1711 and WD 1202-232 have a strong correlation with the best fit binary model, but W0146 and WD 2105-82 are not as strongly correlated. This may be due to higher order systematic noise, which can be improved with kernel phase calibration (see Section \ref{sec:disc}). 

\section{Discussion}\label{sec:disc}
We achieve contrasts of $5-6$ mag for most of the MEOW targets at $\sim200$ mas. The WD distances in this sample range from $6.21-16.47$pc, with an average distance of 12au. The angular resolution of 200 mas (which is within $\lambda/\mathrm{D}$ for \textit{JWST} at $7.7 \mu$m; $\sim244$ mas) at these distances thus translates to $1.52-4.02$ au. Previous analysis of WD 2149+021 using standard imaging on \textit{JWST}/MIRI led to a mass sensitivity of $\sim0.64$M$_\mathrm{_J}$ at the inner most working angle of 654 mas (corresponding to a contrast of at 6.5 mag at 5.8 $\mu$m) \cite{poulsen_MIRI}. Assuming similar performance as presented here, observing this system with KPI would result in a slightly lower mass sensitivity (reaching 6 mag of contrast compared to 6.5 mag), at an inner working angle of $\sim200$mas, within $\lambda/\mathrm{D}$. KPI's mass sensitivity at these separations can help probe the region where inward migrating post-MS giants are thought to reside \cite{unpacking, WDPlanetCandidate, SED_transit_conf_and_MEOW, MEOW_detection}. We expect this sensitivity to improve when robust calibration strategies are applied. 

We successfully recover known companions in WD and BD systems with uncalibrated KPs, with separations ranging from $\sim150$ mas to $\sim2.1$$^{\prime\prime}$ corresponding to $(0.47-4.5) \lambda/\mathrm{D}$. The KPI parameters for three of the companions are in agreement with the DI parameters previously reported \cite{BD_mathew, DIexoplanets}. Likely due to the use of uncalibrated data, the parameters for W0146 (while significant compared to the null model) were discrepant with the DI analysis. Calibration has been shown to decrease the scatter of KPs by removing residual systematic errors\cite{StephAndy, jens_niriss}, which may be required for robust recovery of close-in ($<\lambda/\mathrm{D}$) companions like the one orbiting W0146. Despite these systematic offsets for the closest separations, these companion recoveries demonstrate that even uncalibrated KPI can enable \textit{JWST} super-resolution. They also highlight the prospects for future calibrated KPI MIRI investigations.

Improvements to the MIRI pupil model and exploration of calibration options would benefit MIRI KPI and will be the subject of future work. Adjusting the pupil sampling pitch and transmission cutoff would optimize the pupil model. Decreasing the pupil sampling pitch ($s$) would increase the density of the Fourier phase sampling. We plan to explore various pitch settings to find the minimum viable $s$ value, determined by the $s$ where adjacent (u,v) sampling points no longer add new information. For transmission cutoff, previous KPI studies on \textit{JWST}/NIRISS used a value of 70\% \cite{jens_niriss}, and future work will determine the best cutoff for MIRI. The methodology in Martinache et al.~(2020)\cite{AperatureModeling} can be used to optimize both pitch and transmission cutoff. The expected diagonal redundancy matrix can be compared to the amplitude of the complex visibilities extracted from the data, and they should match for an optimized pupil.

Future work will also determine robust kernel phase observational strategies and calibration methods. Ideally, RDI KPI datasets would include PSF calibrator observations obtained close in time to the science target observations, on the same region of the detector. While testing the limits of KPI calibration may require a dedicated MIRI program, archival datasets with many objects observed close in time could be used to test calibration quality as a function of relative detector position, peak counts, and wavefront drift. Similar comparisons could be performed using multiple PSFs in the same field as the science target. Careful dedicated and archival calibration tests will inform future KPI proposal planning and re-analysis of archival data that was originally processed with traditional methods.
%This may involved looking for archival data or possibly using PSFs in the same field as the science target. With calibrated KPs, we expect to be more sensitive to deeper contrast at a tighter working angle, within half of $\lambda/\mathrm{D}$. This careful calibration will also inform anyone wanting to perform KPI on \textit{JWST}/MIRI archival data that may have been observed without a calibrator.

\section{Conclusion}\label{sec:conc}
We present the first demonstration of KPI on \textit{JWST}/MIRI. We use 16 white dwarfs in the MEOW survey to produce achievable contrast curves, where we are sensitive to companion contrasts of $5-6$ magnitudes for most of the targets at $\sim200$ mas. This angular separation is within $\lambda/\mathrm{D}$ on \textit{JWST} at $7.7 \mu$m, and corresponds to an orbital separation of $\sim2.5$ au, given the average MEOW target distance. Additionally, we are able to recover known companion signals in WD systems and BD systems, with separations of $\sim150-2140$ mas and contrasts of $\sim0.74-4.45$ mag. 

This KPI demonstration uses raw KPs, as well as a pupil sampling pitch chosen by visually inspecting the (u,v) coverage and a transmission cutoff chosen from previous KPI work done on a different \textit{JWST} instrument. We expect the KPI sensitivity to improve with KP calibration, as well as with optimization of the interferometric model pitch and transmission cutoff. Even without calibration, the achievable contrasts presented here reach giant planet masses for WD and BD hosts. This demonstrates the utility of raw KPI for substellar companion searches, motivating its use for re-processing archival data originally processed with traditional methods. Improvements enabled by future work will access even fainter and thus colder / lower-mass companions, making \textit{JWST}/MIRI KPI a promising technique for expanding the directly-imaged exoplanet parameter space.

\acknowledgments % equivalent to \section*{ACKNOWLEDGMENTS}       
 
We would like to thank Cal Bridge for providing financial support for C.A. graduate student research. M.D.F. is supported by an NSF Astronomy and Astrophysics Postdoctoral Fellowship under award AST-2303911.

This work is based on observations made with the NASA/ESA/CSA James Webb Space Telescope. The data were obtained from the Mikulski Archive for Space Telescopes at the Space Telescope Science Institute, which is operated by the Association of Universities for Research in Astronomy, Inc., under NASA contract NAS 5-03127 for JWST. These observations are associated with programs 4403, 2124, and 1911.

All of the data presented in this paper were obtained from the Mikulski Archive for Space Telescopes (MAST) at the Space Telescope Science Institute. The specific observations analyzed can be accessed via DOI \href{https://doi.org/10.17909/zcr4-9k86}{10.17909/zcr4-9k86}. STScI is operated by the Association of Universities for Research in Astronomy, Inc., under NASA contract NAS5–26555. Support to MAST for these data is provided by the NASA Office of Space Science via grant NAG5–7584 and by other grants and contracts.

% References
\bibliographystyle{spiebib} % makes bibtex use spiebib.bst
\bibliography{report} % bibliography data in report.bib

\end{document}